\documentclass[RNAAS]{aastex63}

\newcommand{\N}[1]{{\color{blue}\bf  #1}}

\begin{document}

\title[{\it Gaia} Wd1]{{\it \textbf{Gaia}} EDR3 confirms that Westerlund 1 is closer and older than previously thought}

\correspondingauthor{Mojgan Aghakhanloo}
\email{aghakhanloo@arizona.edu}

\author[0000-0001-8341-3940]{Mojgan Aghakhanloo}
\affiliation{Steward Observatory, University of Arizona, 933 N. Cherry Ave., Tucson, AZ 85721, USA}

\author{Jeremiah W. Murphy}
\affiliation{Department of Physics, Florida State University, 77 Chieftan
Way, Tallahassee, FL 32306, USA}

\author[0000-0001-5510-2424]{Nathan Smith}
\affiliation{Steward Observatory, University of Arizona, 933 N. Cherry Ave., Tucson, AZ 85721, USA}

\author{John Parejko}
\affiliation{Department of Astronomy, University of Washington, Box 351580, Seattle, WA 98195, USA}

\author{Mariangelly D\'iaz-Rodr\'iguez}
\affiliation{Department of Physics, Florida State University, 77 Chieftan
Way, Tallahassee, FL 32306, USA}

\author{Maria R. Drout}
\affiliation{The 
Observatories of the Carnegie Institution for Science, 813 Santa Barbara St, Pasadena}

\author{Jose H. Groh}
\affiliation{School of Physics, Trinity College Dublin, The University of Dublin, Dublin, Ireland}

\author{Joseph Guzman}
\affiliation{Department of Physics, Florida State University, 77 Chieftan
Way, Tallahassee, FL 32306, USA}

\author{Keivan G. Stassun}
\affiliation{Department of Physics \& Astronomy, Vanderbilt University, 6301 Stevenson Center Lane,
Nashville, TN 37235, USA}
\affiliation{Department of
Physics, Fisk University, 1000 17th Avenue N., Nashville, TN 37208, USA}

\begin{abstract}
Using {\it Gaia} Early Data Release 3 (EDR3) parallaxes and Bayesian inference, we infer a parallax of the Westerlund 1 (Wd1) cluster. We find a parallax of $0.34\pm{0.05}$ mas corresponding to a distance of $2.8^{+0.7}_{-0.6}$ kpc. The new {\it Gaia} EDR3 distance is consistent with our previous result using {\it Gaia} DR2 parallaxes.  This confirms that Wd1 is less massive and older than previously assumed. Compared to DR2, the EDR3 individual parallax uncertainties for each star decreased by 30\%.  However, the aggregate parallax uncertainty for the cluster remained the same.  This suggests that the uncertainty is dominated by systematics, which is possibly due to crowding, motions within the cluster, or motions due to binary orbits.
\end{abstract}


\section{} 
Westerlund 1 (Wd1) has previously been discussed as potentially one of the most massive young star clusters in the Galaxy.  Wd1 is of significant interest because it contains a large population of evolved massive stars such as Wolf-Rayet stars, red and blue supergiants, yellow hypergiants, an LBV, and a magnetar \citep{CN04,C05,M06,C06,groh06,F18}. Previous distance estimates to Wd1 ranged from 1.0 to 5.5 kpc \citep{W61,W68,P98,C05,C06}, although values around 5 kpc have usually been adopted. Stellar luminosities at $\sim$5 kpc imply that the cluster's current turnoff mass would be around 40 M$_{\odot}$ or more.

In \cite{A20}, we used {\it Gaia} Data Release 2 \citep[DR2;][]{G16,Ga18} and Bayesian inference to estimate the distance to Wd1. We modeled both cluster stars and Galactic field stars and inferred a parallax of $0.35^{+0.07}_{-0.06}$ mas corresponding to a distance of $2.6^{+0.6}_{-0.4}$ kpc. At this closer distance, stellar luminosities would be reduced by a factor of more than 3.  The turnoff mass would be reduced from $\sim$40 M$_{\odot}$ to around 22 M$_{\odot}$, with a corresponding increase in age and a decrease in the cluster's total stellar mass compared to values usually adopted in the literature.

\begin{figure}[ht!]
\begin{center}
\includegraphics[scale=0.5,angle=0]{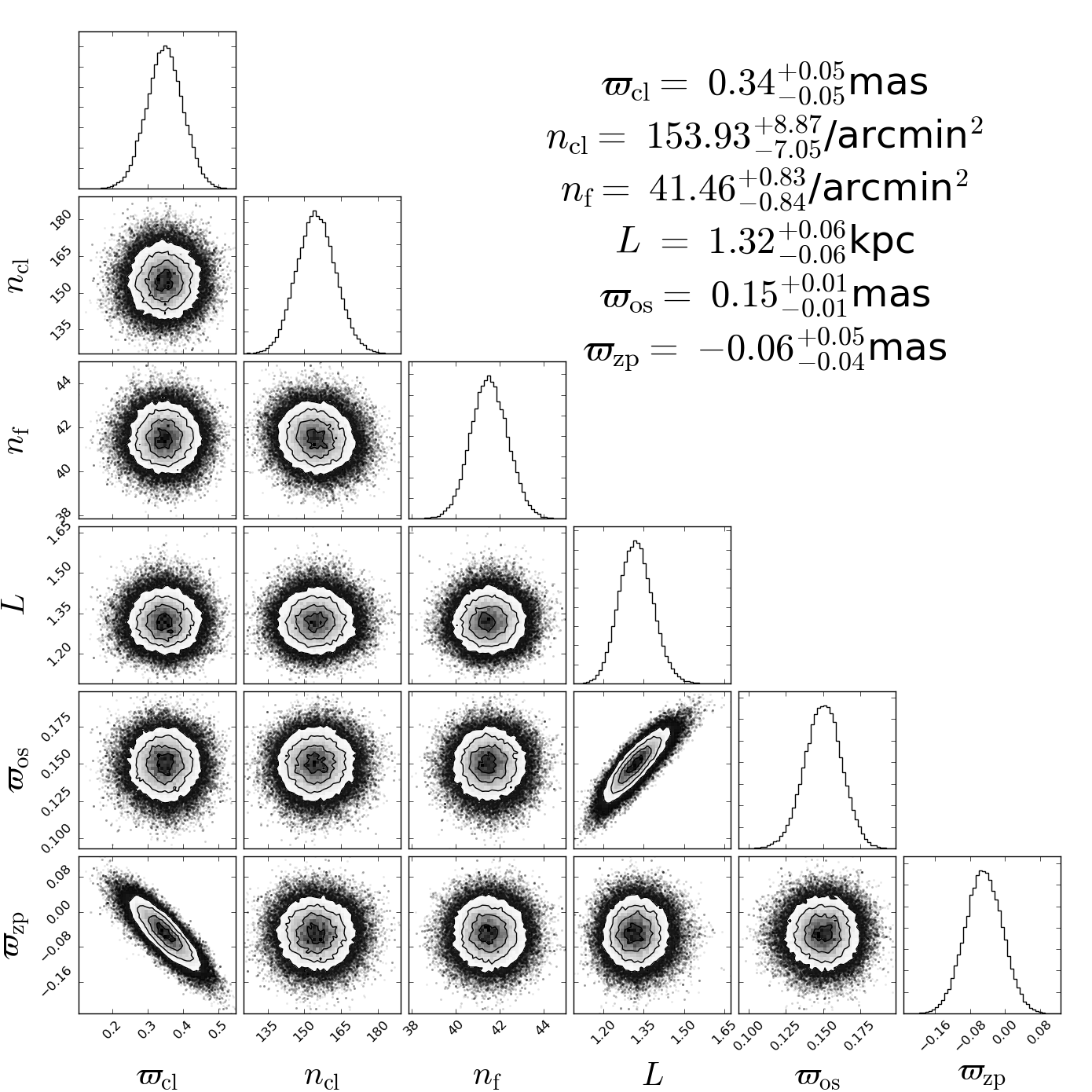}
\caption{Posterior distribution for the six-parameter model. We report the mode and the highest density 68\% confidence interval for the cluster parallax ($\varpi_{\text{cl}}$), the cluster density ($n_{\text{cl}}$), the field-star density ($n_{\text{f}}$), the field-star length scale ($L$), the field-star offset ($\varpi_{\text{os}}$), and the parallax zero-point of the cluster ($\varpi_{\text{zp}}$). The parallax of the cluster is $\varpi_{\text{cl}}=0.34\pm{0.05}$ mas, which corresponds to a distance of $R=2.8^{+0.7}_{-0.6}$ kpc. }\label{fig:Corner}
\end{center}
\end{figure}

In this work, we update a parallax of the cluster using {\it Gaia} early third data release \citep[EDR3;][]{g20}. We infer a parallax of $0.34\pm{0.05}$ mas corresponding to a distance of $2.8^{+0.7}_{-0.6}$ kpc. Fig.~\ref{fig:Corner} shows the posterior distribution for cluster parallax, $\varpi_{\text{cl}}$ (mas), density of the cluster stars, $n_{\text{cl}}$ (number per square arcminute), density of the field stars, $n_{\text{f}}$ (number per square arcminute), the field-star length scale, $L$ (kpc), the field-star offset, $\varpi_{\text{os}}$ (mas), and the parallax zero-point of the cluster, $\varpi_{\text{zp}}$ (mas). The two regions used to constrain these parameters are an inner circle centred on the position of Wd1 and with a radius of 1 arcmin, and an outer annulus from 9 to 10 arcmin.  The values in the top right corner show the mode and the highest 68\% density interval (HDI) for marginalized distributions. The density of the cluster is $n_{\text{cl}}=153.93^{+8.87}_{-7.05}$ stars per square arcminute, density of field stars is $n_{\text{f}}=41.46^{+0.83}_{-0.84}$ stars per square arcminute, the field-star length scale is $L=1.32\pm{0.06}$ kpc, the field-star offset is $\varpi_{\rm{os}}=0.15\pm{0.01}$ mas, and the parallax zero-point of the cluster is $\varpi_{\text{zp}}=-0.06^{+0.05}_{-0.04}$ mas.

The new {\it Gaia} EDR3 result is consistent with our previous work using {\it Gaia} DR2. In the {\it Gaia} EDR3, the individual parallax errors decreased by 30\% \citep{g20}. Also, in this sample, the number of sources in the inner circle with a good solution (at least eight visibility periods, RUWE $<1.40$ and astrometric excess noise sigma $\leq$ 2) increased by a factor of $\sim$3. Even though the individual {\it Gaia} EDR3 parallax precision for  each  star increased, the new {\it Gaia} EDR3 parallax of the Wd1 cluster has the same precision due to unmodeled systematic errors. If the uncertainty is dominated by random statistics, then the uncertainty should be of order $\sigma_i/\sqrt{N}$, where $N$ is the number of stars with good solutions and $\sigma_i$ is the uncertainty of each star.  Therefore, if uncertainties are random, {\it Gaia} EDR3 parallax uncertainty of Wd1 cluster should be a factor of $\sim$2 smaller than {\it Gaia} DR2 parallax uncertainty.  The fact that the {\it Gaia} EDR3 parallax precision of the cluster stays the same implies that there is a systematic error that is unmodeled.  Such systematic errors could be due to crowding, motions within the cluster or motions due to binary orbits. Due to the increased number of sources in the inner circle, the cluster density increases by a factor of $\sim$1.5. The {\it Gaia} EDR3 field-star length scale is within $\sim$1.6 sigma of the {\it Gaia} DR2 result. The field-star offset and the parallax zero-point of the cluster are consistent with the previous results using {\it Gaia} DR2. 

W243 is a confirmed Luminous Blue Variable (LBV) that is associated with the Wd1 cluster \citep{C04,C05}. In {\it Gaia} DR2, the parallax of the individual star W243 is $0.979 \pm{0.165}$ mas, implying a distance of 1.78$^{+2.37}_{-0.95}$~kpc \citep{smith19}, while the {\it Gaia} EDR3 parallax of the individual star W243 is $0.012\pm{0.081}$ mas. In both {\it Gaia} DR2 and EDR3, the excess astrometric noise sigma for this star is larger than 2, which indicates that the source may not be astrometrically well-behaved. The significant difference between {\it Gaia} DR2 and EDR3 data and large astrometric excess noise sigma may be due to crowding, and binarity in this region. Therefore, the distance to the Wd1 cluster is a more reliable distance estimate to LBV W243 than its individual distance estimate, and the cluster parallax also provides a much more precise estimate of the distance.

While we infer that possible systematic effects seem to limit the improvement in precision in the parallax of the Wd1 cluster as we move from {\it Gaia} DR2 to EDR3, we nevertheless find a result that is consistent with our previously inferred distance.  This confirms that the Wd1 cluster is in fact closer, less massive, and less luminous than typically assumed in the literature, having the important consequence that the magnetar, the LBV, and other evolved stars seen in the cluster descended from initial masses far less than 40 M$_{\odot}$, being closer to 25 M$_{\odot}$ or less.

\bibliographystyle{aasjournal}
\bibliography{MA}

\begin{thebibliography}{}
\expandafter\ifx\csname natexlab\endcsname\relax\def\natexlab#1{#1}\fi
\providecommand{\url}[1]{\href{#1}{#1}}
\providecommand{\dodoi}[1]{doi:~\href{http://doi.org/#1}{\nolinkurl{#1}}}
\providecommand{\doeprint}[1]{\href{http://ascl.net/#1}{\nolinkurl{http://ascl.net/#1}}}
\providecommand{\doarXiv}[1]{\href{https://arxiv.org/abs/#1}{\nolinkurl{https://arxiv.org/abs/#1}}}

\bibitem[{{Aghakhanloo} {et~al.}(2020){Aghakhanloo}, {Murphy}, {Smith},
  {Parejko}, {D{\'\i}az-Rodr{\'\i}guez}, {Drout}, {Groh}, {Guzman}, \&
  {Stassun}}]{A20}
{Aghakhanloo}, M., {Murphy}, J.~W., {Smith}, N., {et~al.} 2020, \mnras, 492,
  2497, \dodoi{10.1093/mnras/stz3628}

\bibitem[{Brown {et~al.}(2018)Brown, Vallenari, Prusti, de~Bruijne, Babusiaux,
  Bailer-Jones, Biermann, Evans, Eyer, \& et~al.}]{Ga18}
Brown, A. G.~A., Vallenari, A., Prusti, T., {et~al.} 2018, A\&A, 616, A1,
  \dodoi{10.1051/0004-6361/201833051}

\bibitem[{Clark \& Negueruela(2003)}]{CN04}
Clark, J.~S., \& Negueruela, I. 2003, A\&A, 413, L15,
  \dodoi{10.1051/0004-6361:20031700}

\bibitem[{Clark {et~al.}(2005)Clark, Negueruela, Crowther, \& Goodwin}]{C05}
Clark, J.~S., Negueruela, I., Crowther, P.~A., \& Goodwin, S.~P. 2005, A\&A,
  434, 949, \dodoi{10.1051/0004-6361:20042413}

\bibitem[{{Clark, J. S.} {et~al.}(2005){Clark, J. S.}, {Larionov, V. M.}, \&
  {Arkharov, A.}}]{CL05}
{Clark, J. S.}, {Larionov, V. M.}, \& {Arkharov, A.} 2005, A\&A, 435, 239,
  \dodoi{10.1051/0004-6361:20042563}

\bibitem[{{Clark, J. S.} \& {Negueruela, I.}(2004)}]{C04}
{Clark, J. S.}, \& {Negueruela, I.} 2004, A\&A, 413, L15,
  \dodoi{10.1051/0004-6361:20031700}

\bibitem[{Collaboration {et~al.}(2020)Collaboration, Brown, Vallenari, Prusti,
  de~Bruijne, Babusiaux, \& Biermann}]{g20}
Collaboration, G., Brown, A. G.~A., Vallenari, A., {et~al.} 2020, Gaia Early
  Data Release 3: Summary of the contents and survey properties.
\newblock \doarXiv{2012.01533}

\bibitem[{Crowther {et~al.}(2006)Crowther, Hadfield, Clark, Negueruela, \&
  Vacca}]{C06}
Crowther, P.~A., Hadfield, L.~J., Clark, J.~S., Negueruela, I., \& Vacca, W.~D.
  2006, \mnras, 372, 1407, \dodoi{10.1111/j.1365-2966.2006.10952.x}

\bibitem[{Fenech {et~al.}(2018)Fenech, Clark, Prinja, Dougherty, Najarro,
  Negueruela, Richards, Ritchie, \& Andrews}]{F18}
Fenech, D.~M., Clark, J.~S., Prinja, R.~K., {et~al.} 2018, A\&A, 617, A137,
  \dodoi{10.1051/0004-6361/201832754}

\bibitem[{Groh {et~al.}(2006)Groh, Damineli, Teodoro, \& Barbosa}]{groh06}
Groh, J.~H., Damineli, A., Teodoro, M., \& Barbosa, C.~L. 2006, A\&A, 457, 591,
  \dodoi{10.1051/0004-6361:20064929}

\bibitem[{Muno {et~al.}(2005)Muno, Clark, Crowther, Dougherty, de~Grijs, Law,
  McMillan, Morris, Negueruela, Pooley, \& et~al.}]{M06}
Muno, M.~P., Clark, J.~S., Crowther, P.~A., {et~al.} 2005, ApJ, 636, L41,
  \dodoi{10.1086/499776}

\bibitem[{Piatti {et~al.}(1998)Piatti, Bica, \& Clari{\'a}}]{P98}
Piatti, A.~E., Bica, E., \& Clari{\'a}, J.~J. 1998, A\&AS, 127, 423,
  \dodoi{10.1051/aas:1998111}

\bibitem[{Prusti {et~al.}(2016)Prusti, de~Bruijne, Brown, Vallenari, Babusiaux,
  Bailer-Jones, Bastian, Biermann, Evans, \& et~al.}]{G16}
Prusti, T., de~Bruijne, J. H.~J., Brown, A. G.~A., {et~al.} 2016, A\&A, 595,
  A1, \dodoi{10.1051/0004-6361/201629272}

\bibitem[{{Smith} {et~al.}(2019){Smith}, {Aghakhanloo}, {Murphy}, {Drout},
  {Stassun}, \& {Groh}}]{smith19}
{Smith}, N., {Aghakhanloo}, M., {Murphy}, J.~W., {et~al.} 2019, \mnras, 488,
  1760, \dodoi{10.1093/mnras/stz1712}

\bibitem[{Westerlund(1961)}]{W61}
Westerlund, B.~E. 1961, PASP, 73, 51, \dodoi{10.1086/127618}

\bibitem[{Westerlund(1968)}]{W68}
---. 1968, ApJ, 154, L67, \dodoi{10.1086/180270}

\end{thebibliography}
\end{document}